\documentclass[11pt]{article}
\usepackage{appendix}
\usepackage{multicol}
\usepackage{calc}
\usepackage{enumitem}
\usepackage[margin=0.5in]{geometry}
\usepackage{graphicx}
\usepackage{caption}
\SetLabelAlign{parright}{\parbox[t]{\labelwidth}{\raggedleft#1}}
\DeclareGraphicsExtensions{.png,.jpg}

\begin{document}

\title{Stochastic Spot/Volatility Correlation in Stochastic Volatility Models and Barrier Option Pricing}
\author{Mark Higgins \\
Washington Square Technologies}

\maketitle

\begin{abstract}
Most models for barrier pricing are designed to let a market maker tune the model-implied covariance between moves in the asset spot price and moves in the implied volatility skew. This is often implemented with a local volatility/stochastic volatility mixture model, where the mixture parameter tunes that covariance. This paper defines an alternate model where the spot/volatility correlation is a separate mean-reverting stochastic variable which is itself correlated with spot. We also develop an efficient approximation for barrier option and one touch pricing in the model based on semi-static vega replication and compare it with Monte Carlo pricing. The approximation works well in markets where the risk neutral drift is modest.
\end{abstract}

\section{Introduction}

Barrier option pricing is sensitive to parameters of stochastic volatility models mostly through what the model implies about the covariance between the asset spot price and the implied volatility skew (see Appendix \ref{bardepcorr} for an elaboration of this intuition). In most markets there is indeed significant positive realized correlation between spot and volatility skew, and this dynamic can materially impact fair prices for barrier options.

In foreign exchange markets, where barrier options are a liquid ``vanilla exotic" product, this is generally implemented with some flavor of local volatility/stochastic volatility (LVSV) mixture model (for example, \cite{lvsv}). A typical formulation is:

\begin{eqnarray}
\frac{dS}{S} &=& \mu \, dt + \sigma(S,t) \sqrt{v} \, dw_s \nonumber \\
dv &=& \beta (\overline{v}-v) dt + \alpha \sqrt{v} \, dw_v \nonumber \\
<dw_s \, dw_v> &=& \rho \, dt \nonumber 
\end{eqnarray}

Here, $S$ is the asset spot price and $v$ represents a stochastic factor on the instantaneous variance that follows a mean-reverting square root process. $\mu$ is the normal risk-neutral drift of spot, equal to the discount rate $r$ minus the asset discount rate $q$.

In this LVSV model the local volatility term $\sigma(S,t)$ is calibrated to to implied volatilities and ``fills in" whatever match to implied volatilities is not handled by the pure stochastic volatility piece. This construction means that the volatility of volatility parameter $\alpha$ effectively controls the mixture between local and stochastic volatility, and therefore the amount that implied volatility skew moves when the asset spot price moves. In the limit where $\alpha$ goes to zero, LVSV is a pure local volatility model, where the model's implied volatility skew moves maximally as the spot price moves. In the limit where $\alpha$ and $\rho$ are set such that, with constant $\sigma$, the model most closely matches the market implied volatilities, LVSV's implied volatility skew moves the least as spot moves. 

Therefore, in LVSV, the $\alpha$ parameter controls the model-implied covariance between spot and implied volatility skew. However, it suffers from the same key problem as a pure local volatilty model: the contribution to forward skew and smile from the local volatility piece tend to zero as time progresses, which generally causes the model to misrepresent the real future market state.

Another approach is one by Carr\cite{carr07}, where skew is modeled by jumps in spot and the frequencies of up- and down-jumps are dependent on the level of spot. This model has the benefit of quick vanilla pricing, which lets one calibrate the model quickly to market implied volatilities, an important feature of any model. It can give a stationary term structure of skew and smile, so can have more realistic forward volatilities than LVSV. However it often requires jumps in spot that are much larger than those actually experienced, which can misprice barrier option pricing if the probability of spot jumping through the barrier is significant.

This paper develops a different approach to incorporating the covariance between spot and implied volatility skew. We start by recognizing that, in stochastic volatility models, implied volatility skew is closely related to (and usually close to linear in) the model correlation between spot and instantaneous volatility.

Therefore we model this spot/volatility correlation as a separate stochastic asset, one that is correlated with spot. The correlation between spot and spot/volatility correlation, combined with volatility of spot/volatility correlation, is what determines the covariance between spot and implied volatility skew, and is the aspect of the model dynamics which most significantly impacts barrier option pricing. 

The end goal is a fast approximation for barrier option prices that is parameterized by only vanilla option prices and a small number of extra marked parameters that determine the term structure of the covariance between spot and implied volatility skew. This approximation will be motivated by and compared against pricing in a formal model.

The full specification of the formal model that we use as a benchmark is:

\begin{eqnarray}
\frac{dS}{S} &=& \mu \, dt + \sqrt{v} \, dw_s \nonumber \\
dv &=& \beta (\overline{v}-v) dt + \alpha \sqrt{v} \, dw_v \nonumber \\
<dw_s \, dw_v> &=& \rho \, dt \nonumber \\
d\rho &=& \gamma (\overline{\rho} - \rho) dt + \epsilon \, \sqrt{1 - \rho^2} \, \sqrt{v} \, dw_{\rho} \nonumber \\
<dw_s \, dw_{\rho}> &=& \rho_{cs} \, dt \nonumber \\
<dw_v \, dw_{\rho}> &=& \rho \, \rho_{cs} \, dt
\label{eqsdes}
\end{eqnarray}

where $S$ is the asset spot price, $v$ is the instantaneous variance (volatility squared), and $\rho$ is the spot/volatility correlation, as per the Heston\cite{heston93} stochastic volatility model. $\rho$, however, is generalized to be stochastic and follows its own mean-reverting process.

We name this benchmark model the stochastic volatility/stochastic correlation, or SVSC, model.

The nine SVSC model parameters are:

\begin{description}[labelwidth=\widthof{\bfseries aaaaaa},align=parright]
\item[$\beta$] The mean reversion strength of instantaneous variance
\item[$\overline{v}$] The long-run average instantaneous variance
\item[$v(0)$] The initial value of instantaneous variance
\item[$\alpha$] The volatility of volatility
\item[$\gamma$] The mean reversion strength of spot/volatility correlation
\item[$\overline{\rho}$] The long-run average spot/volatility correlation
\item[$\rho(0)$] The initial value of spot/volatility correlation
\item[$\epsilon$] The volatility of spot/volatility correlation
\item[$\rho_{cs}$] The correlation of FX spot and the spot/volatility correlation
\end{description}

Note that the correlation between instantaneous variance and spot/volatility correlation is set to $\rho \, \rho_{cs}$ and is not a separate model parameter. This choice maximizes the determinant of the 3x3 correlation matrix and forces that matrix to be always positive definite. Most derivative prices are only weakly dependent on this correlation, so we set it for maximal convenience.

\begin{figure}[h]
\includegraphics[width=\textwidth]{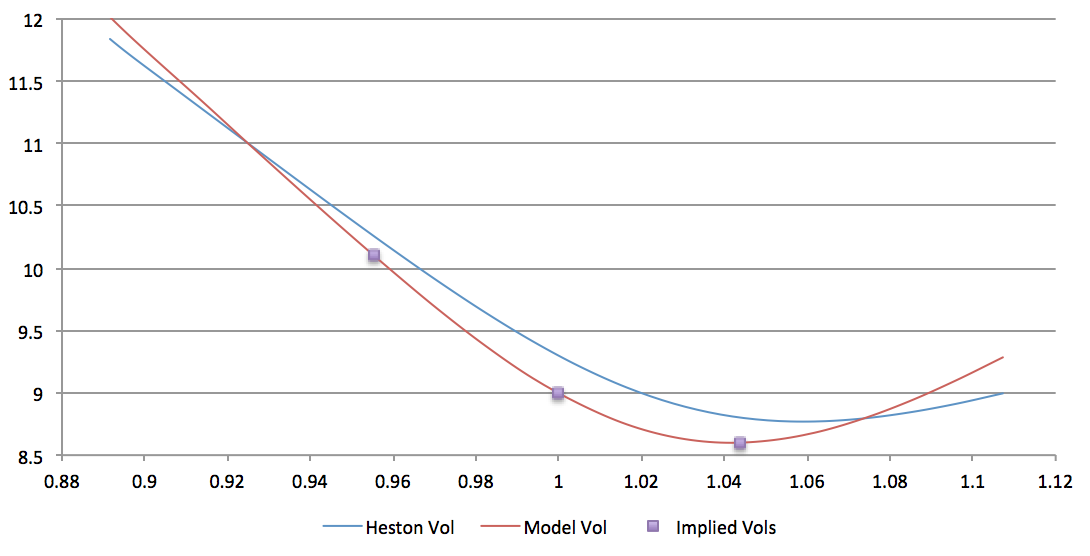}
\caption{Implied volatilities of the SVSC model, after calibrating $\overline{v}=v(0)$, $\overline{\rho}=\rho(0)$, and $\alpha$ to three market implied volatilities. Heston model implied volatilities are also displayed - note that SVSC has larger volatility smile than Heston because of the extra contribution from stochastic spot/volatility correlation.}
\label{figvols}
\end{figure}

Figure \ref{figvols} shows an example implied volatility vs strike chart for parameters representative of a G10 foreign exchange market. A priori SVSC parameters were $\beta=2$, $\gamma=4$, $\epsilon=10$, and $\rho_{cs}=0.7$. $\overline{v}$, $\alpha$, and $\overline{\rho}$ are calibrated to the market implied volatilities at the 25-delta and at-the-money strikes, and $v(0)$ and $\rho(0)$ are assumed to equal their long-run average values. Also displayed in the same figure are the implied volatilities for the Heston\cite{heston93} model - assuming that $\epsilon=0$ but otherwise using the calibrated SVSC parameters. The implied volatilities we calibrate to are 10.1\% for strike 0.9554; 9.0\% for strike 1.0000; and 8.6\% for strike 1.0438 (the forward is equal to 1 in this example). The calibrated parameters are $\overline{v}=v(0)=(9.962\%)^2$, $\alpha=0.2536$, and $\overline{\rho}=\rho(0)=-0.3835$.

Unlike LVSV, SVSC has approximately stationary forward skew and smile, and unlike Carr\cite{carr07}'s model, SVSC has no jumps in spot.

\section{An Approximation to Barrier Option Pricing}

Exotic derivative pricing generally involves taking market prices for forwards and implied volatilities as inputs; calibrating a formal model to the market data; and then using that calibrated model to price exotics consistently with prices of the vanilla hedges.

A model like the benchmark SVSC model above, with three stochastic factors, requires a Monte Carlo or backward induction numerical scheme to price even vanilla options. These numerical schemes are generally quite slow, especially when calibrating model parameters to market implied volatilities.

We develop an approximation for barrier option pricing that starts with a semi-static replication for the barriers: a replication that is a good static vega hedge for the barrier, but only as long as the barrier is not touched. If the barrier is touched the replication hedge needs to be ``unwound": that is, closed out at then-current market prices. The approximation involves deciding on the replicating portfolio for a given barrier option and then estimating the expected cost of unwinding the replication hedge if the barrier is touched.

We calculate the replication cost, unwind premium, and barrier price in the Heston model (with an adjusted correlation parameter in unwind to incorporate the SVSC effects), then divide the barrier price by the Heston-model price of the underlying vanilla option to get the no touch probability in a measure associated with the underlying vanilla. The final approximation price of the barrier option equals the market price\footnote{In this paper the ``market'' price is taken to be the SVSC price of the vanilla to ease comparisons with the formal model, but in practice it would come from a standard vanilla interplation scheme.} (not the Heston-model price) of the vanilla option multiplied by that no touch probability. This final step is performed to ensure that the approximation does not suffer from inaccuracies due to slight differences in implied volatility interpolation between models. 

One key advantage of this approach is that no complex model calibrations to market implied volatilities are required. The approximation will require a calibration of a Heston model to market implied volatilities at the barrier option expiration date, but Heston has a semi-closed form for vanilla pricing and that calibration is very fast.

In the following we will assume barrier options are knockout options. Knockin options of course can be priced using in-out parity, where a knockout plus the equivalent knockin equals the underlying vanilla option.

\subsection{Replication and Unwind Approach}

Normally when calculating the expected cost of unwinding the replication hedge we would need to integrate over the three-dimensional probability distribution of first touching the barrier at time t conditioned on instantaneous volatility $v$ and spot/volatility correlation $\rho$ when the barrier is touched.

Our replication approach is chosen to simplify this calculation dramatically. First we choose a replication that is a good global vega hedge so long as the barrier is not touched, which collapses the integration in the $v$ direction. Second, we choose a replication whose price is roughly linear in $\rho$ when spot is at the barrier, which means that we can collapse the integration in the $\rho$ direction and simply replace it by its expected value conditioned on hitting the barrier.

\subsection{Heston with Time-Dependent Spot/Volatility Correlation}
\label{sechestoncorr}

We care that the replication is a good vega hedge because that means the price of the replication portfolio is only weakly sensitive on the level of instantaneous variance if spot touches the barrier. In that case the unwind cost mostly depends on the level of spot/volatility correlation at unwind, and therefore on what the model implies for covariance between spot and spot/volatility correlation, or in market terms, the implied volatility skew at that point. In addition the price of the replication portfolios we choose will be roughly linear in the spot/volatility correlation at unwind, so pricing depends only on the expected value of that correlation, not on the rest of its distribution.

That fact lets us use the Heston\cite{heston93} model as a proxy for SVSC when calculating expected unwind prices, but using a time-dependent but deterministic spot/volatility correlation. The Heston formulation allows for (almost) closed form pricing only with constant spot/volatility correlation; to allow us to use the standard result we next develop an approximation for an ``effective" constant spot/volatility correlation that can be plugged into the standard Heston result that replicates the value in the generalized Heston model with a time-dependent determinist correlation of the form

\begin{equation}
\rho(t) = \overline{\rho} + (\rho(0)-\overline{\rho}) e^{-\gamma t}
\label{rhot}
\end{equation}

which is of course the expected value of $\rho(t)$ in the formal model from Equation \ref{eqsdes}.

In foreign exchange markets implied volatility skew is parameterized through the ``risk reversal", normally defined as the difference in implied volatility between a call option with Black-Scholes delta equal to 0.25 and a put option with a Black-Scholes delta of -0.25\footnote{More precisely that is the ``25-delta risk reversal" - there are risk reversal values for any delta less than 50, but the most liquid is the 25-delta risk reversal, and we will simply call that the ``risk reversal".}.

Consider an option position that is closely related to the risk reversal implied volatility difference defined above: a ``risk reversal position'' is an option portfolio that is long the 25-delta call option and short the 25-delta put option. This portfolio has a price which is close to linear in the risk reversal implied volatility difference.

The expected vanna at time $t$ of a risk reversal position initiated at time 0, $\langle V(t) \rangle$, is approximately proportional to $\frac{(T-t)}{\sigma T}$, where $T$ is the initial time to expiration and $\sigma$ is the constant Black-Scholes volatility (see Appendix \ref{secvds}). This observation lets us estimate the price premium in the Heston model of the risk reversal position expiring at time $T$ over its Black-Scholes value, $v_{RR}(T)$, which is mostly due to cross gamma between spot and implied volatility:

\begin{eqnarray}
v_{RR}(T) &\approx& \int_{t=0}^T{\langle V(t) \rangle \rho \, \alpha_I \, \sigma \rangle dt} \\
          &\approx& \int_{t=0}^T{A \frac{(T-t)}{\sigma T} \rho(t) \alpha \frac{(1-e^{-\beta (T-t)})}{\beta (T-t)} \sigma \, dt} \\
          &=& \frac{A \alpha}{\beta T} \int_{t=0}^T{\rho(t) (1-e^{-\beta (T-t)}) dt}
\end{eqnarray}

where $A \frac{(T-t)}{\sigma T}$ is the approximate expected vanna ($A$ is a function of the initial delta of the risk reversal), $\sigma_I$ is an assumed level of implied volatility (roughly equal to instantaneous volatility $\sigma$, and assuming a flat term structure of ATM volatility), $\alpha_I \sqrt{dt}$ is the standard deviation of the move in implied volatility over time $dt$, $\alpha$ is the volatility of volatility parameter in the Heston model, and $\beta$ is the mean reversion speed of volatility in the Heston model. $\rho(t)$ is the (deterministic) instantaneous spot/volatility correlation at time $t$. Note that this assumes zero contribution from ``volga'' (second order exposure to volatility) as a risk reversal position has close to zero volga at initiation and close to zero on average over its life.

Substituting $\rho(t)$ from equation \ref{rhot}, we find that the premium of the risk reversal position over Black-Scholes is approximately:

\begin{equation}
v_{RR}(T) \approx \frac{A \alpha}{\beta} \left( \overline{\rho} \, D_1(T) + (\rho(0)-\overline{\rho}) \, D_2(T) \right)
\label{eqrrprem}
\end{equation}

where $D_1(T)$ and $D_2(T)$ are functions of $\beta$, $\gamma$, and time to expiration $T$:

\begin{eqnarray}
D_1(T) &=& 1 - \frac{(1-e^{-\beta T})}{\beta T} \\
D_2(T) &=& \frac{(1-e^{-\gamma T})}{\gamma T} + \frac{(e^{-\beta T}-e^{-\gamma T})}{(\beta-\gamma)T}
\end{eqnarray}

We want an approximate constant spot/volatility correlation $\rho_a$ which reproduces the same risk reversal premium as our time-dependent correlation term structure, so we equate the risk reversal premium with constant correlation $\rho_a$ with that in Equation \ref{eqrrprem}:

\begin{equation}
\rho_a(T) = \overline{\rho} + (\rho(0)-\overline{\rho}) D_2(T)/D_1(T)
\label{eqrhoa}
\end{equation}

\subsection{Out-of-the-Money Barrier Options}
\label{secotmbar}

An ``out-of-the-money barrier option" is one where the underlying vanilla option is out of the money vs spot when the barrier is hit - for example, a down-and-out call.

\subsubsection{Out-of-the-Money Barrier Option Replication}

A good semi-static replication for an out-of-the-money barrier option involves two vanilla options: long one unit of the vanilla option underlying the barrier option; and short a vanilla option with a strike ``reflected" through the barrier (the ``reflected strike" $K'$). The short option is a put if the barrier option is a call, and a call if the barrier option is a put. 

The quantity of the short option in the replication is $\sqrt{K/K'}$. This quantity is not simply 1 because of a constraint that if we flip the numeraire from the denominated currency to the asset and regenerate the replication, we should get the same replicating portfolio. This symmetry under switch of numeraire is important in the foreign exchange space where there is no preferred numeraire, but is good practice generally and improves the quality of the vega replication.

The strike of the short option - the reflected strike $K'$ - is set such that the initial price of the replication under the Black-Scholes model matches the initial price of the barrier option under Black-Scholes:

\begin{equation}
v_{R,BS}(K') \sqrt{\frac{K}{K'}} = v_{0,BS}(K) - v_{B,BS}(B,K)
\end{equation}

where $v_{R,BS}(K')$ is the Black-Scholes price of the reflected-strike vanilla option (whose call/put type is opposite to that of the barrier option); $v_{0,BS}(K)$ is the Black-Scholes price of the vanilla option with strike $K$ underlying the barrier option; and $v_{B,BS}(B,K)$ is the Black-Scholes price of the barrier option with strike $K$ and barrier level $B$. ``Black-Scholes price" here means Black-Scholes with constant volatility equal to the at-the-money volatility to the barrier option expiration date $T$ and constant risk neutral drift $\mu$. This equation must be solved numerically but, since all the Black-Scholes prices have closed form, the root-finding is very fast.

If the risk neutral drift $\mu=0$, the reflected strike $K'$ reduces to $K'=B^2/K$. For non-zero $\mu$ it is similar to that value but approximately corrects for the fact that the forward will be different than the barrier level if spot hits the barrier at a time $t<T$.

This replication follows, for example, Carr, Ellis, and Gupta\cite{carr98}, where the replication quantities are not functions of volatility. That means the replication ends up being a good semi-static vega hedge for the barrier option (only semi-static, not static, because the hedge fails if spot touches the barrier).

\begin{figure}[h]
\includegraphics[width=\textwidth]{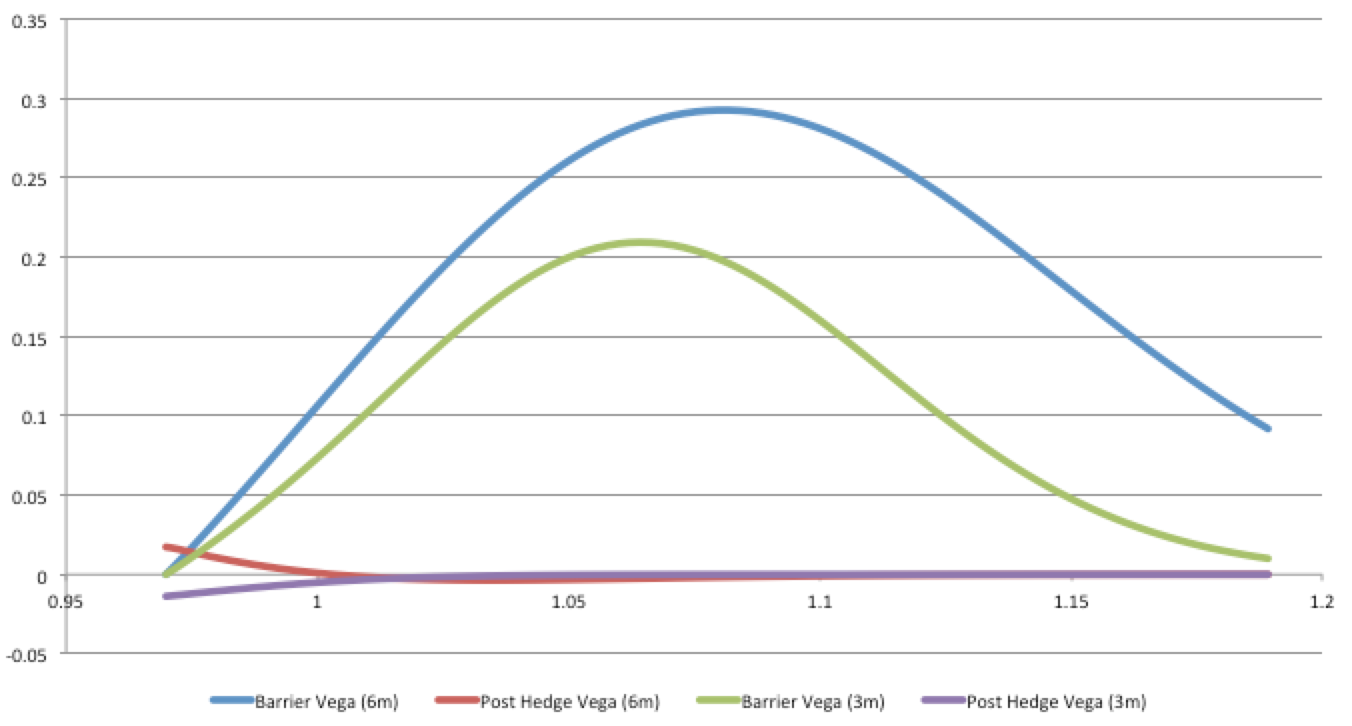}
\caption{Black-Scholes vega for a down-and-out call option, strike 1 and barrier 0.97, before and after hedging with the vanilla replication. Vega is shown for two times: once with six months to expiration, and once with three months to expiration. The replication was set based on prices with spot=1 and six months to expiration.}
\label{figbarvega5}
\end{figure}

The effectiveness of the vega hedge is displayed in Figure \ref{figbarvega5}, which shows the Black-Scholes vega of a down-and-out call option with strike 1 and barrier level 0.97. The market corresponds to that in Figure \ref{figvols}, except with a risk neutral drift of -5\%. Even with a substantial risk neutral drift the vega hedge is a good semi-static vega hedge across different spots and times. The corresponding chart for zero risk neutral drift is not shown because the post-hedge vega is indistinguishable from zero.

\subsubsection{Out-of-the-Money Barrier Option Unwind}

In our approximation we assume the price of the barrier option $v_B$ is given as:

\begin{equation}
v_B = v_R(0,S(0)) - v_U
\end{equation}

where $v_R(t,S)$ is the price of the vanilla replication portfolio as seen at time $t$ and spot $S$, and $v_U$ is the expected cost of unwinding it due to touching the barrier at different times before expiration $T$:

\begin{equation}
v_U = \int_{t=0}^T{\langle v_R(t,B) \rangle D(t) p(t) dt}
\end{equation}

Here, $D(t)$ is the denominated currency discount factor from time 0 to time $t$, and $p(t) dt$ is the probability of first touching the barrier in $t \rightarrow t+dt$. $\langle v_R(t,B) \rangle$ is the expected value of the replication portfolio as seen at time $t$ conditioned on spot being at the barrier $B$.

We approximate that integral with a sum over $N$ equally-space time intervals:

\begin{equation}
v_U \approx \sum_{i=0}^N{\langle v_R(t_{mi},B) \rangle D(t_{mi}) ( P(t_{ei}) - P(t_{si}) )}
\label{eqotmunwind}
\end{equation}

where $t_{si}$ is the time to the start of the $i$\textsuperscript{th} interval ($t_{s0}=0$), $t_{mi}$ is the time to the middle of the $i$\textsuperscript{th} interval, and $t_{ei}$ is the time to the end of the $i$\textsuperscript{th} interval ($t_{eN}=T$, and $t_{ei}=t_{s(i+1)}$). $P(t)$ is the probability of first touch before time $t$, equal to the price of a one touch expiring at time $t$ with no discounting.

An important property of the replication portfolio is that, because it looks like a risk reversal position when spot is at the barrier, its price is relatively insensitive to the level of instantaneous volatility; and because a risk reversal position's value is roughly linear in spot/volatility correlation, its price depends only on the expected value of spot/volatility correlation, not its variance.

Therefore we write:

\begin{equation}
\langle v_R(t,B) \rangle \approx v_{RH}(t,\beta,\rho_{aH}(t,B),\alpha_H,v_H(t),\overline{v_H})
\end{equation}

where $v_{RH}(t,\beta,\rho_H,\alpha_H,v_H(t),\overline{v_H})$ represents the price of the replication portfolio under the Heston model, as seen at time $t$. The Heston model parameters are the mean reversion strength $\beta$ of instantaneous variance, the constant spot/volatility correlation $\rho_H$, the volatility of volatility $\alpha_H$, the instantaneous variance $v_H(t)$, and the long term variance $\overline{v_H}$.

We will estimate the appropriate Heston-model parameters to use from a calibration of the Heston model at $t=0$ to vanilla option prices expiring at time $T$. $\beta$ is specified externally, and in the calibration we assume $\overline{v_H}=v_H(0)$, so there are three parameters to calibrate at the start.

For the instantaneous variance parameter $v_H(t)$ we use the Dupire\cite{dupire94} volatility, and using Heston-model option prices to time $t$ based on the initial calibration to vanilla options expiring at time $T$.

The final parameter to estimate, and the most important, is the value for the Heston spot/volatility correlation parameter $\rho_H$. For this we will estimate the expected value of the instantaneous spot/vol correlation, conditioned on spot hitting the barrier at time $t$, and use Equation \ref{eqrhoa} to get the effective constant Heston correlation parameter $\rho_H$.

We need two additional approximations here: first, that the conditional expected correlation does not depend much on the level of volatility, so we can use an approximate constant volatility; and second, that the stochastic term in the SDE for spot/volatility correlation $\rho$ does not depend on $\rho$. That is, we simplify our system of SDEs to:

\begin{eqnarray}
dx &=& (\mu-\frac{\sigma^2}{2}) \, dt + \sigma \, dw_s \nonumber \\
d\rho &=& \gamma (\overline{\rho} - \rho) dt + \epsilon' \sigma \, dw_{\rho} \nonumber \\
\langle dw_s dw_\rho \rangle &=& \rho_{cs} \, dt
\end{eqnarray}

where $\sigma$ is some approximately constant volatility and $\epsilon' = \epsilon \sqrt{1-\rho'^2}$, with $\rho'$ some approximate constant correlation (more on that at the end of this section). $x=\ln{S}$, the log of spot.

Now we replace $dw_\rho$ with a $\rho_{cs} dw_s + \sqrt{1-\rho^2} dw_0$, where $dw_0$ has zero correlation with $dw_s$, and replace $dw_s$ with its equivalent in terms of $d\ln{S}$:

\begin{equation}
\label{eqdrho}
d\rho = \gamma (\overline{\rho} - \rho) dt + \rho_{cs} \epsilon' ( dx - \mu \, dt + \frac{\sigma^2}{2} dt ) + \sqrt{1-\rho_{cs}^2} \, \epsilon' \sigma dw_0
\end{equation}

We want to condition on $S(t)=B$ for some time $t$ in the future when spot first touches the barrier $B$, so we replace the SDE for $x$ with a Brownian bridge process:

\begin{equation}
dx(s) = \frac{(x(t)-x(s))}{(t-s)} dt + \sigma dz_s
\end{equation}

The expected value of $x(s)$ following a Brownian bridge process is just $x(t) s/t$, and the expected value of $dx(s)$ is $x(t)/t$.

We can then take the expected value of Equation \ref{eqdrho} to get

\begin{equation}
\frac{d \langle \rho \rangle}{ds} = \gamma (\overline{\rho} - \langle \rho \rangle) + \rho_{cs} \epsilon' ( \frac{x(t)}{t} - \mu + \frac{\sigma^2}{2} )
\end{equation}

This ODE solves to

\begin{equation}
\langle \rho(t) \rangle = \overline{\rho} + ( \rho(0) - \overline{\rho} ) e^{-\gamma t} + \rho_{cs} \epsilon' \frac{(1-e^{-\gamma t})}{\gamma t} ( x(t) - (\mu-\frac{\sigma^2}{2}) t )
\end{equation}

The first two terms give the unconditional expected value of $\rho(t)$, and the third gives the correction from conditioning on the final value of spot.

For our purpose, $x(t)=\ln \left( B/S(0) \right)$, where $B$ is the barrier level and $S(0)$ is the initial level of spot. For convenience we also set $\rho(0)=\overline{\rho}=\rho_H$, where $\rho_H$ is the correlation parameter from an initial Heston calibration (performed at $t=0$) to vanillas expiring at time $T$.

Writing the initial forward to $t$ as $F(0,t)=S(0) e^{\mu t}$, we get the final expression for conditional expected instantaneous spot/volatility correlation:

\begin{equation}
\langle \rho(t) \rangle = \rho_H + \rho_{cs} \epsilon' \frac{(1-e^{-\gamma t})}{\gamma t} \left( \ln{\left( \frac{B}{F(0,t)} \right) } + \frac{\sigma^2 t}{2} \right)
\end{equation}

Note that the dependence on the specific volatility choice $\sigma$ is indeed somewhat weak: it appears only in the final term, and $\ln \left( B/F(0,t) \right)$ is order $\sigma \sqrt{t}$, much larger than $\sigma^2 t/2$ in most cases. For $\sigma^2$ we use the value of $\overline{v}_H$ from the calibration at $t=0$ to vanillas expiring at $T$.

Earlier we left out what the best value is to use for $\rho'$ in the definition of $\epsilon'$. An algorithm that works quite well is to make two passes: first use $\rho_H$ and get an initial estimate of $\rho_1 = \langle \rho(t) \rangle$ from that; then refine $\rho'=(\rho_H+\rho_1)/2$ and get the final estimate of $\langle \rho(t) \rangle$ using that approximate average correlation over the move to the barrier.

We use that second estimate of conditional expected instantaneous spot/volatility correlation to get the Heston spot/volatility correlation parameter $\rho_{aH}$ to use in the unwind calculation as:

\begin{equation}
\rho_{aH}(t,B) = \overline{\rho} + ( \langle \rho(t) \rangle - \overline{\rho} ) D_2(T-t)/D_1(T-t)
\label{eqrhoah}
\end{equation}

With this we have all the parameters we need to calculate the expected conditional cost of unwinding the replication portfolio if the barrier is first touched at time $t$. But to calculate the full expected unwind cost we also need to estimate $P(t)$, the probability of first touching the barrier by time $t$.

The unwind value is not that sensitive to this probability; instead the result depends mostly on the predictions for conditional spot/volatility correlation. So we do not need to be too accurate here. One approach is to use the Black-Scholes one touch pricing formula, but we can do somewhat better than that by again looking at an approximate one touch replication.

A reasonable replication for a one touch is a European digital with the same strike as the one touch, but twice the notional: when spot is at the barrier the one touch has value 1 and the European digital has a price of roughly 50\%, as it has (roughly) even odds of ending in the money or out of the money. We will use a relatively crude approximation for the one touch price based on that replication:

\begin{equation}
P(t) = P_{BS}(t) + 2 (E(t)-E_{BS}(t)) (1-P_{BS}(t))
\end{equation}

where $P_{BS}(t)$ is the Black-Scholes price of the one touch; $E(t)$ is the Heston-model price of the European digital; and $E_{BS}(t)$ is the Black-Scholes price of the European digital (using the at-the-money implied volatility as the constant Black-Scholes volatility). All prices are divided by discount factors to turn them into probabilities.

This lets us calculate the expected cost of unwinding the replication in Equation \ref{eqotmunwind}.

\subsubsection{Comparison with Monte Carlo}

Barrier option prices in G10 foreign exchange markets have very tight bid/offer spreads, which means that model accuracy is very important. Typical bid/offer spreads for out-of-the-money barriers are only slightly wider than those of vanilla options, usually on the order of a few basis points.

Table \ref{tabotmbarsrf0} shows results for the same set of example market data as Figure \ref{figvols}. Spot is equal to 1 and risk neutral drift is zero, as is the discount rate. Model and Heston prices in the examples were calculated with Monte Carlo simulation using 1,000 time steps and 1,000,000 paths to get standard errors down to approximately 0.2bp. The unwind calculation used $N=10$ time buckets. The approximation does a very good job of matching the formal model price for out-of-the-money barrier options, in almost all cases landing within 1bp of the benchmark SVSC model price. Differences between the full model and Heston prices for barrier options are usually a few basis points, and the accuracy of the approximation is better than that difference.

\begin{center}
\begin{tabular}{| c | c | c | c | c | c | c | c |}
\hline
\textbf{Call/} & \textbf{Strike} & \textbf{Barrier} & \textbf{Model} & \textbf{Approx} & \textbf{Approx} & \textbf{Heston} & \textbf{BS} \\
\textbf{Put} & & & \textbf{Price} & \textbf{Price} & \textbf{Diff} & \textbf{Diff} & \textbf{Diff} \\ \hline
Call & 0.95 & 0.900 & 589.2 & 589.5 & -0.3 & +1.3 & +16.7 \\ \hline
Call & 1.00 & 0.900 & 251.3 & 251.2 & +0.1 & +0.0 & -2.5 \\ \hline
Call & 1.00 & 0.950 & 226.8 & 226.0 & +0.8 & +3.6 & -12.7 \\ \hline
Call & 1.00 & 0.975 & 167.2 & 166.8 & +0.4 & +3.1 & -9.6 \\ \hline
Call & 1.01 & 0.980 & 119.7 & 119.5 & +0.2 & +3.5 & -10.5 \\ \hline
Call & 1.03 & 0.990 & 46.7 & 46.5 & +0.2 & +2.9 & -7.9 \\ \hline
Call & 1.10 & 0.950 & 19.6 & 18.8 & +0.8 & +1.9 & -2.2 \\ \hline
Put  & 0.90 & 1.050 & 36.1 & 36.7 & -0.6 & +1.4 & +23.9 \\ \hline
Put  & 0.97 & 1.010 & 54.5 & 54.1 & +0.4 & +2.0 & +2.9 \\ \hline
Put  & 0.99 & 1.020 & 126.9 & 125.8 & \textbf{+1.1} & +2.5 & +0.0 \\ \hline
Put  & 1.00 & 1.025 & 172.0 & 107.8 & \textbf{+1.2} & +2.5 & -2.1 \\ \hline
Put  & 1.00 & 1.050 & 230.9 & 230.9 & +0.0 & +2.3 & -5.7 \\ \hline
Put  & 1.00 & 1.100 & 251.8 & 252.1 & -0.3 & -0.8 & -1.8 \\ \hline
Put  & 1.05 & 1.100 & 568.7 & 569.3 & -0.6 & +0.2 & -11.4 \\ \hline
\end{tabular}
\captionof{table}{Barrier price comparison for out-of-the-money barrier options when risk neutral drift $\mu=0$. All prices and price differences are displayed in basis points. Approx Diff is the difference between the formal model price and the approximation; Heston Diff is the difference between the model price and the Heston price of the barrier option; and BS Diff is the difference between the model price and the Black-Scholes price of the barrier option. Approximation errors greater than 1bp are highlighted in bold.}
\label{tabotmbarsrf0}
\end{center}

The approximation is less accurate when risk neutral drift is significant. Table \ref{tabotmbarsrf5} shows results for the same market data except with a risk neutral drift of -5\%. Approximation error in the case of an up-and-out put option with strike 0.97 and barrier level 1.01 is over 2bp, which is comparable to the bid/ask spread. In most cases however the approximation error is 1bp or less.

\begin{center}
\begin{tabular}{| c | c | c | c | c | c | c | c |}
\hline
\textbf{Call/} & \textbf{Strike} & \textbf{Barrier} & \textbf{Model} & \textbf{Approx} & \textbf{Approx} & \textbf{Heston} & \textbf{BS} \\
\textbf{Put} & & & \textbf{Price} & \textbf{Price} & \textbf{Diff} & \textbf{Diff} & \textbf{Diff} \\ \hline
Call & 0.95 & 0.900 & 397.5 & 398.6 & \textbf{-1.1} & +0.9 & +7.3 \\ \hline
Call & 1.00 & 0.900 & 135.6 & 136.0 & -0.4 & -0.2 & -10.7 \\ \hline
Call & 1.00 & 0.950 & 119.6 & 120.1 & -0.5 & +3.4 & -16.8 \\ \hline
Call & 1.00 & 0.975 & 84.7 & 85.6 & -0.9 & +4.2 & -12.3 \\ \hline
Call & 1.01 & 0.980 & 57.4 & 58.0 & -0.6 & +4.1 & -11.2 \\ \hline
Call & 1.03 & 0.990 & 20.3 & 19.4 & +0.9 & +2.6 & -6.1 \\ \hline
Call & 1.10 & 0.950 & 9.7 & 9.4 & +0.3 & +1.6 & +2.3 \\ \hline
Put  & 0.90 & 1.050 & 55.5 & 55.8 & -0.4 & +1.1 & +26.4 \\ \hline
Put  & 0.97 & 1.010 & 93.0 & 95.1 & \textbf{-2.1} & +1.0 & -1.9 \\ \hline
Put  & 0.99 & 1.020 & 210.6 & 210.9 & -0.3 & +1.1 & -4.2 \\ \hline
Put  & 1.00 & 1.025 & 279.7 & 280.0 & -0.3 & +0.3 & -4.1 \\ \hline
Put  & 1.00 & 1.050 & 359.6 & 358.9 & +0.7 & +1.1 & -11.9 \\ \hline
Put  & 1.00 & 1.100 & 382.6 & 382.8 & -0.2 & -0.7 & -10.4 \\ \hline
Put  & 1.05 & 1.100 & 777.7 & 778.5 & -0.8 & +0.4 & -5.8 \\ \hline
\end{tabular}
\captionof{table}{Barrier price comparison for out-of-the-money barrier options when risk neutral drift $\mu=-5\%$. All prices and price differences are displayed in basis points. Approx Diff is the difference between the formal model price and the approximation; Heston Diff is the difference between the model price and the Heston price of the barrier option; and BS Diff is the difference between the model price and the Black-Scholes price of the barrier option. Approximation errors greater than 1bp are highlighted in bold.}
\label{tabotmbarsrf5}
\end{center}

\subsection{One Touches}
\label{secot}

In-the-money, or reverse, barrier options knock out when the underlying vanilla option is in the money, and the replication described in Section \ref{secitmbarrepl} requires pricing of one touches. Before we develop our approximation for in-the-money barrier options, then, we need to develop one for one touches. Here, a one touch is defined as a derivative that, if a barrier is touched at any time before the expiration $T$, pays one unit of the denominated currency at $T$. That is, payment is always at expiration, not at time of touch.

\subsubsection{One Touch Replication}

The replication we choose for a one touch is twice the notional of a European digital and the same strike. This replication acts as an effective semi-static vega hedge as long as the risk neutral drift is modest; large risk neutral drift means that the European digital price is significantly different from 50\% when spot is at the barrier, and twice the European digital does not have a price at that boundary approximating the unit payoff of the one touch.

\begin{figure}[h]
\includegraphics[width=\textwidth]{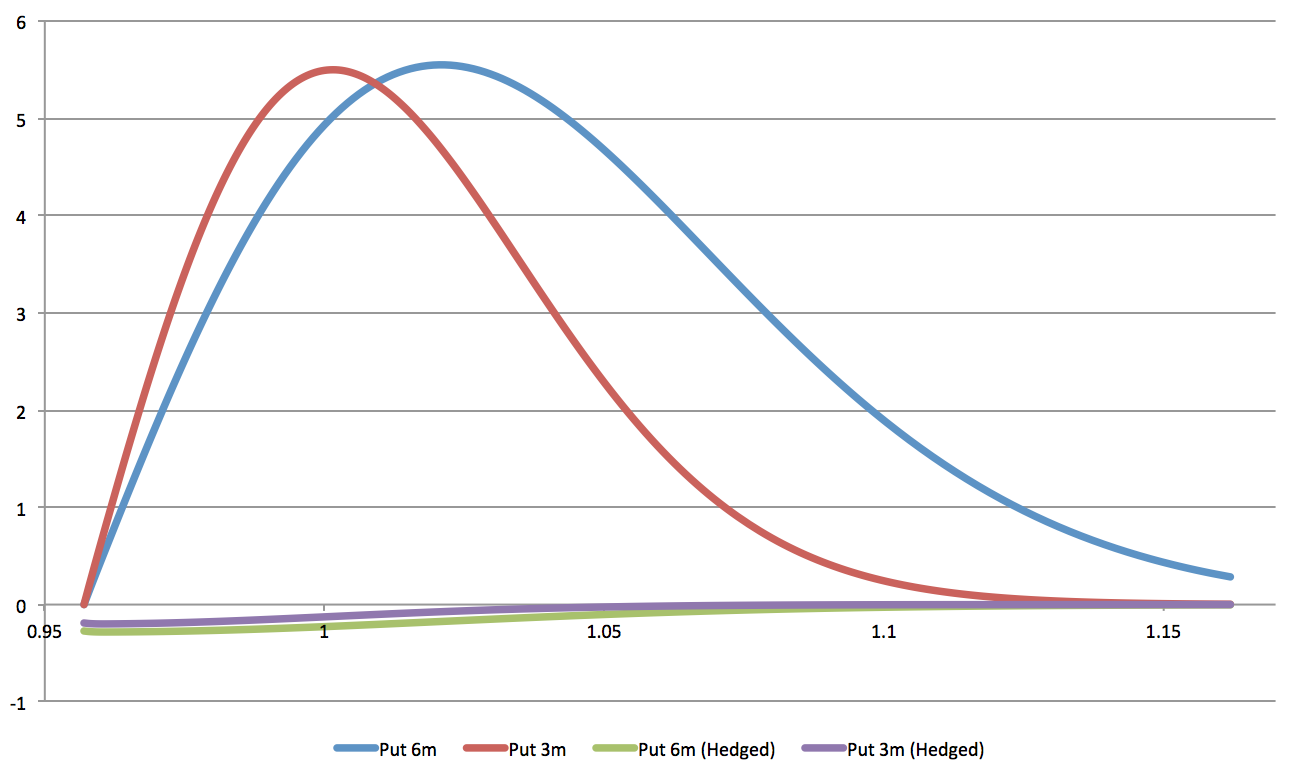}
\caption{Black-Scholes vega for a one touch ``put'' that knocks down and in at 0.9569, which has an initial price of 50\% at spot=1 and initial time to expiration of six months, and in a market with zero risk neutral drift. The vega is shown as a function of spot at two times: at origination with six months to expiration; and after three months, with three months remaining to expiration. Two lines show the vega of the one touch by itself; the other two show the vega after hedging with the European digital replication. The replication is an excellent global vega hedge for the one touch.}
\label{figotvegarf0}
\end{figure}

Figure \ref{figotvegarf0} shows the Black-Scholes vega of a one touch alone and after hedging with this replication as a function of spot when the risk neutral drift is zero. Plots are shown for time to expiration 0.5y, when the one touch and the replication hedge are initiated, as well as time to expiration 0.25y for the same portfolio. At-the-money volatility is assumed to be 9\%, initial spot is 1, and the strike of the one touch is 0.9569, giving it an initial Black-Scholes price of 50\%. In this example the post-hedge vega is always very small compared to the scale of the one touch vega before the hedge.

\begin{figure}[h]
\includegraphics[width=\textwidth]{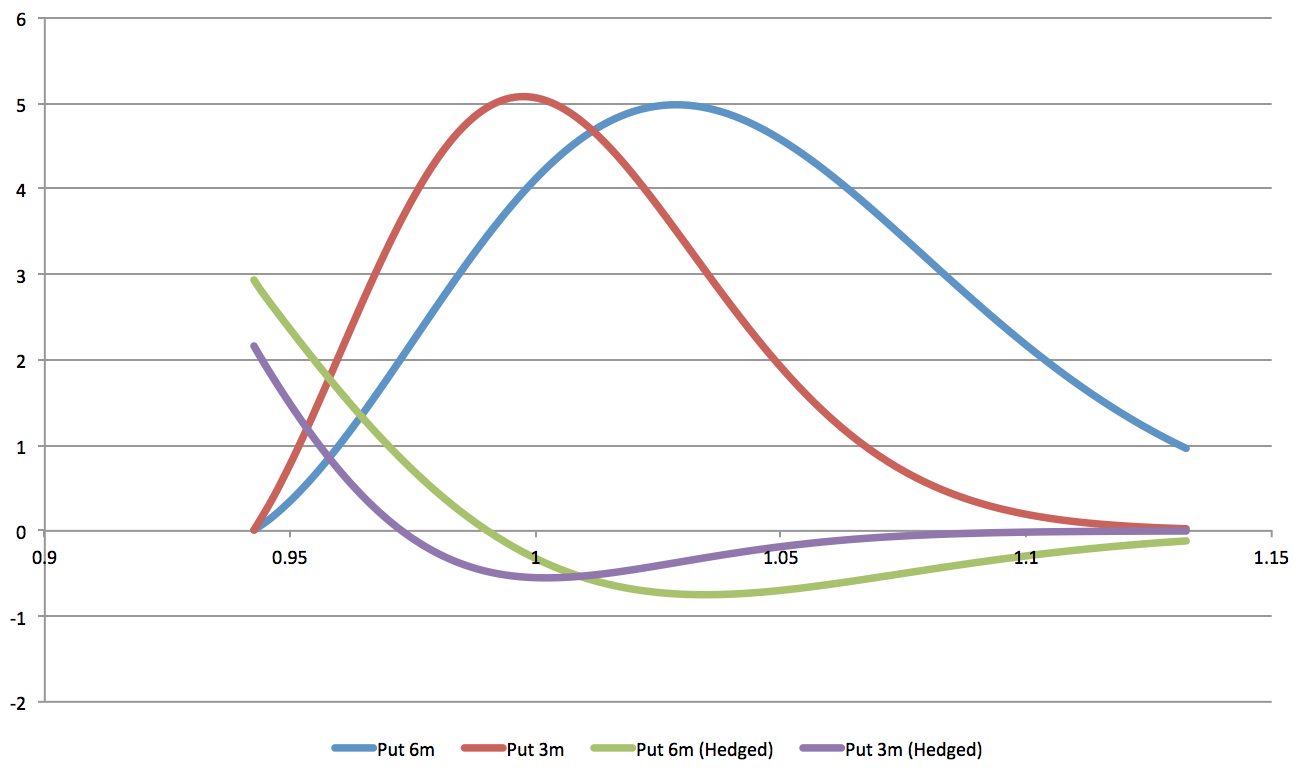}
\caption{Black-Scholes vega for a one touch ``put'' that knocks down and in at 0.9425, which has an initial price of 50\% at spot=1 and initial time to expiration of six months, and in a market with a -5\% risk neutral drift. The vega is shown as a function of spot at two times: at origination with six months to expiration; and after three months, with three months remaining to expiration. Two lines show the vega of the one touch by itself; the other two show the vega after hedging with the European digital replication. The replication is an reasonable global vega hedge for the one touch, though performs somewhat poorly when spot is near the barrier.}
\label{figotvegarf5}
\end{figure}

Figure \ref{figotvegarf5} shows the same thing but for the case where risk neutral drift is -5\%. In this case the replication is a good vega hedge when spot is near its initial value or far from the barrier, but deteriorates as an effective vega hedge if spot approaches the barrier.

\subsubsection{One Touch Unwind}

The unwind calculation for a one touch is similar to that for an out-of-the-money barrier option; that is, like Equation \ref{eqotmunwind}, the unwind value is

\begin{equation}
v_U \approx \sum_{i=0}^N{\langle v_R(t,B) \rangle D(t_{mi}) ( P(t_{ei}) - P(t_{si}) )}
\label{eqotunwind}
\end{equation}

where in this case the expected price of the replication portfolio represents the price of the European digital replication, which is also mostly dependent on the level of the implied volatility skew when spot is at the barrier. As before we will use the Heston price of the European digital at the barrier, using the calibrated level of the Heston volatility of volatility parameter $\alpha_H$ and the Dupire instantaneous volatility, plus the effective spot/volatility correlation $\rho_{aH}(t,B)$ derived in Equation \ref{eqrhoah}:

\begin{equation}
\langle v_R(t,B) \rangle \approx v_{RH}(t,\rho_{aH}(t,B),\alpha_H,v_H(t),\overline{v_H})
\end{equation}

\subsubsection{Comparison with Monte Carlo}

One touch bid/ask spreads in G10 foreign exchange are typically 2-3\%, so approximation errors should be less than that to be effective. The approximation works well in the case of small risk neutral drift and starts to lose its power for larger risk neutral drifts.

\begin{figure}[h]
\includegraphics[width=\textwidth]{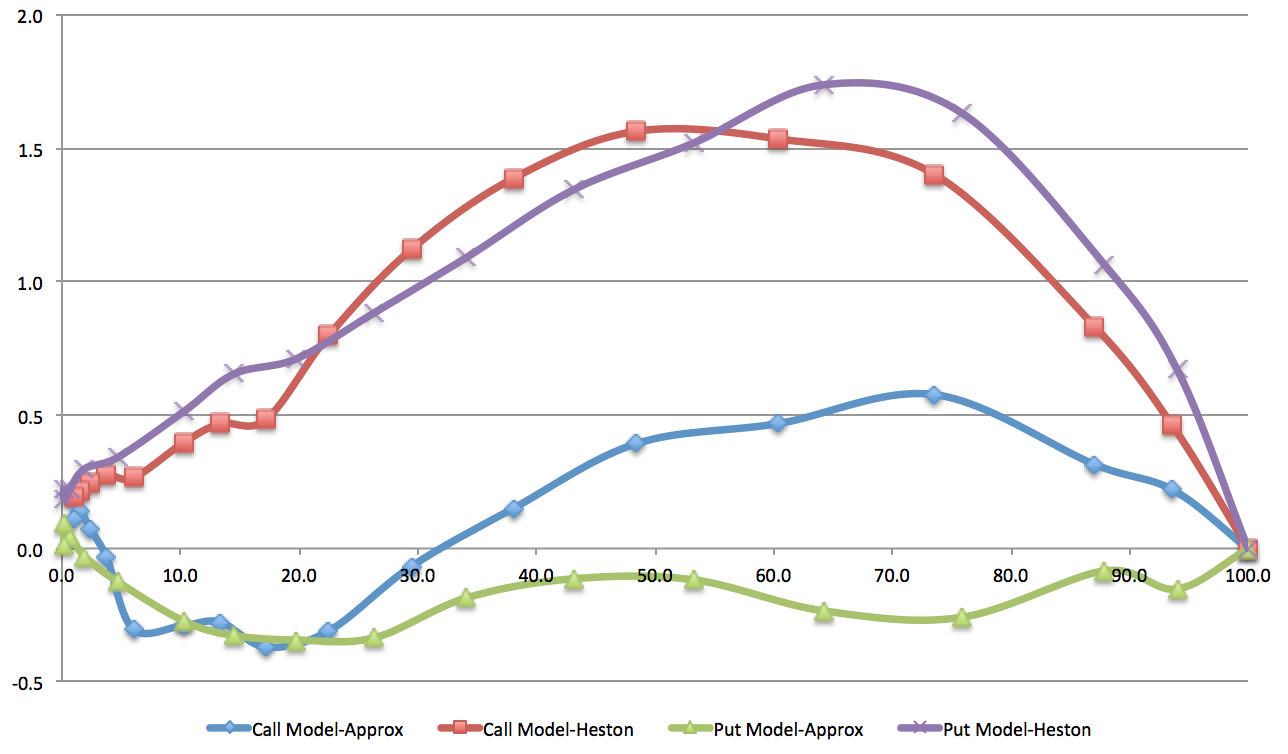}
\caption{Comparison of approximate vs model prices in the case of zero risk neutral drift. The x-axis represents the Black-Scholes price of the one touch. Plots are shown for one touch ``calls'', where the strike is above current spot, and for one touch ``puts'', where strike is below current spot. For each, two plots are included: one that shows the SVSC model price less the approximation price to demonstrate approximation error; and one that shows the SVSC model price less the Heston model price, as we want our approximation to do better than standard Heston in accuracy.}
\label{figotmcrf0}
\end{figure}

Figure \ref{figotmcrf0} shows approximation results for the same set of example market data as Figure \ref{figvols}. Spot is equal to 1 and risk neutral drift is zero, as is the discount rate is zero. Model and Heston prices in the examples were calculated with Monte Carlo simulation using 1,000 time steps and 1,000,000 paths to get standard errors down to approximately 0.1\% in one touch price. The unwind calculation used $N=10$ time buckets. The x-axis is the Black-Scholes price of the one touch. Four plots are displayed: for calls and puts (where a one touch call knocks up and a put knocks down) the difference between the model price and the Heston model price, both calculated with Monte Carlo simulation; and the difference between the difference between the Monte Carlo model price and the approximation price. For both calls and puts the approximation error is mostly less than 0.5\% in price and small compared to the difference between the model price and the Heston model price.

\begin{figure}[h]
\includegraphics[width=\textwidth]{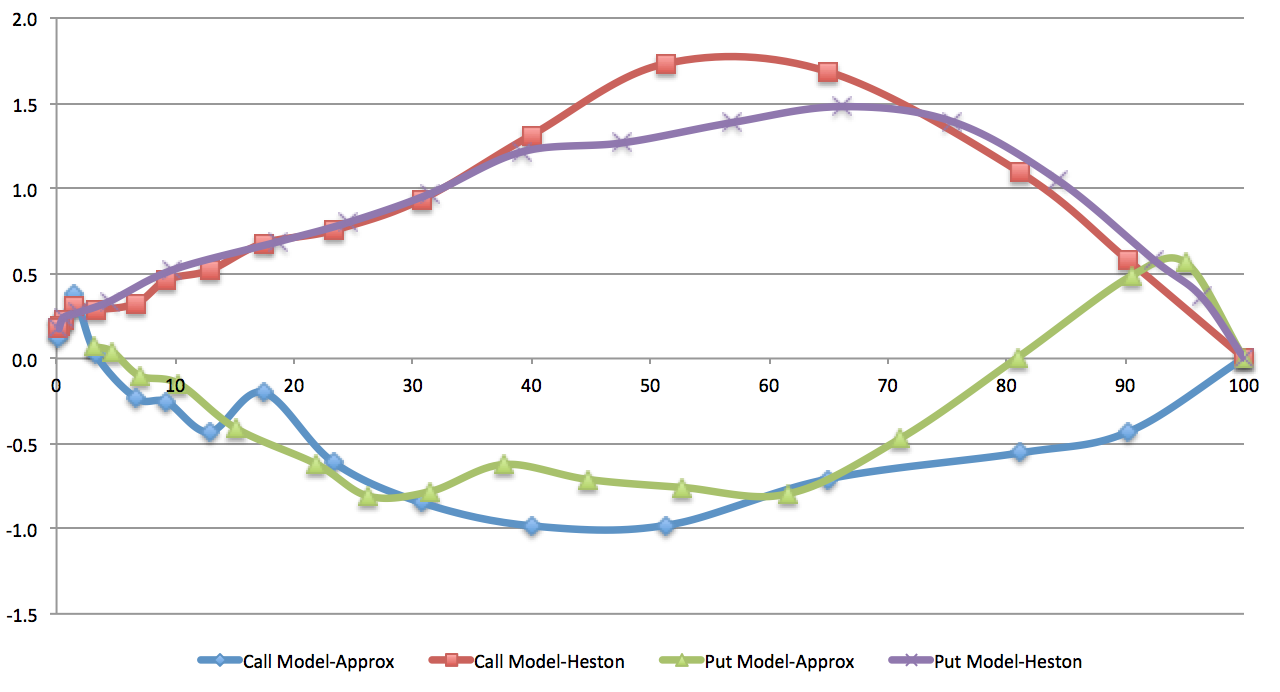}
\caption{Comparison of approximate vs model prices in the case of risk neutral drift equal to -5\%. The x-axis represents the Black-Scholes price of the one touch. Plots are shown for one touch ``calls'', where the strike is above current spot, and for one touch ``puts'', where strike is below current spot. For each, two plots are included: one that shows the SVSC model price less the approximation price to demonstrate approximation error; and one that shows the SVSC model price less the Heston model price, as we want our approximation to do better than standard Heston in accuracy.}
\label{figotmcrf5}
\end{figure}

Figure \ref{figotmcrf5} shows approximation results for the same set of example market data, but in the case where the risk neutral drift is -5\%. The approximation error is noticeably larger than in the case of zero risk neutral drift, but still within 1\% of the model price calculated with Monte Carlo.

\subsection{In-the-Money Barrier Options}

An in-the-money barrier option is one where the option is in the money when spot is at the barrier. 

\subsubsection{In-the-Money Barrier Option Replication}
\label{secitmbarrepl}

We use put/call parity for barrier options to replicate the in-the-money barrier option as an out-of-the-money barrier option and a portfolio of one touches. Barrier put/call parity means

\begin{equation}
v_{bc}(K,B) - v_{bp}(K,B) = v_{bf}(K,B)
\end{equation}

where $v_{bc}(K,B)$ is the price of a barrier call option with strike $K$ and barrier $B$; $v_{bp}(K,B)$ is the price of a barrier put option; and $v_{bf}(K,B)$ is the price of a barrier forward. The barrier forward price can be replicated with a strip of one touches:

\begin{equation}
v_{bf}(K,B) = S e^{-q T} - K e^{-r T} - \int_{t=0}^{T}{ ( B e^{-q (T-t)} - K e^{-r (T-t)} ) e^{-r t} \, p(t) dt }
\end{equation}

where $S$ is current spot, $r$ is the discount rate, $q$ is the asset discount rate, $T$ is the time to expiration, and $p(t) dt$ is the probability of first touching the barrier $B$ in time $t \rightarrow t+dt$. The price of a one touch $v_{ot}(B,T)$ is

\begin{equation}
v_{ot}(B,T) = e^{-r T} \int_{t=0}^{T}{ p(t) dt }
\end{equation}

which lets us integrate by parts to derive the replication of a barrier forward in terms of one touches of different tenors:

\begin{equation}
v_{bf}(K,B) = S e^{-q T} - K e^{-r T} - (B-K) v_{ot}(B,T) + B (q-r) \int_{t=0}^{T}{ e^{-q (T-t)} v_{ot}(B,t) dt }
\end{equation}

Note that if the risk neutral drift $r-q$ is zero, the replication involves just a single one touch to the forward expiration date; otherwise it involves a strip of one touches with expirations from trade initiation to forward expiration.

With the representation of a barrier forward as a strip of one touches we can calculate the price of an in-the-money barrier option in terms of an out-of-the-money barrier option as per Section \ref{secotmbar} and one touches as per Section \ref{secot}.

\subsubsection{Comparison with Monte Carlo}

Bid/ask spreads for in-the-money barrier options are typically made by applying a one touch bid/ask spread to the payoff jump across the barrier level on the expiration date - the difference between the option strike price and the barrier level. Therefore the appropriate comparison for approximation error depends on the payoff. In the results here we assume a 2\% bid/ask spread on the payoff jump.

Table \ref{tabitmbarsrf0} shows results for the same set of example market data as Figure \ref{figvols}. Spot is equal to 1 and risk neutral drift is zero, as is the discount rate. Model and Heston prices in the examples were calculated with Monte Carlo simulation using 1,000 time steps and 1,000,000 paths to get standard errors down to less than 1bp even for the cases where the payoff jump is largest. The approximation errors are significantly larger than for the out-of-the-money barrier option case, but that is due to the significant payoff jump risk; in all cases the approximation error is less than 13\% of the bid/ask spread.

\begin{center}
\begin{tabular}{| c | c | c | c | c | c | c | c | c |}
\hline
\textbf{Call/} & \textbf{Strike} & \textbf{Barrier} & \textbf{Model} & \textbf{Approx} & \textbf{Approx} & \textbf{Heston} & \textbf{BS} & \textbf{Bid/Ask} \\
\textbf{Put} & & & \textbf{Price} & \textbf{Price} & \textbf{Diff} & \textbf{Diff} & \textbf{Diff} & \textbf{Spread} \\ \hline
Call & 0.95 & 1.200 & 573.2 & 578.0 & -4.5 & -4.6 & +9.2 & 50 \\ \hline
Call & 1.00 & 1.200 & 232.6 & 236.3 & -3.6 & -4.1 & -13.6 & 40 \\ \hline
Call & 1.00 & 1.100 & 148.8 & 147.6 & +1.3 & -3.5 & +23.2 & 20 \\ \hline
Call & 1.00 & 1.050 & 40.8 & 41.9 & -1.2 & -4.1 & +20.5 & 10 \\ \hline
Call & 1.01 & 1.100 & 108.3 & 107.2 & +1.1 & -3.0 & +15.7 & 18 \\ \hline
Call & 1.03 & 1.100 & 49.3 & 48.9 & +0.5 & -2.1 & +5.2 & 14 \\ \hline
Call & 1.10 & 1.200 & 10.6 & 12.1 & -1.5 & -1.3 & -5.3 & 20 \\ \hline
Put  & 0.90 & 0.800 & 15.5 & 15.7 & -0.3 & -1.3 & +3.7 & 20 \\ \hline
Put  & 0.97 & 0.900 & 27.9 & 26.6 & +1.3 & -2.0 & -27.7 & 14 \\ \hline
Put  & 0.99 & 0.900 & 64.7 & 62.5 & +2.2 & -3.3 & -48.1 & 18 \\ \hline
Put  & 1.00 & 0.950 & 18.9 & 17.7 & +1.2 & -2.5 & -5.1 & 10 \\ \hline
Put  & 1.00 & 0.900 & 92.1 & 89.5 & +2.6 & -4.0 & -58.6 & 20 \\ \hline
Put  & 1.00 & 0.800 & 208.0 & 208.8 & -0.9 & -4.7 & -44.8 & 40 \\ \hline
Put  & 1.05 & 0.800 & 517.6 & 518.8 & -1.2 & -5.7 & -64.2 & 50 \\ \hline
\end{tabular}
\captionof{table}{Barrier price comparison for in-the-money barrier options when risk neutral drift $\mu=0$. All prices and price differences are displayed in basis points. Approx Diff is the difference between the formal model price and the approximation; Heston Diff is the difference between the model price and the Heston price of the barrier option; BS Diff is the difference between the model price and the Black-Scholes price of the barrier option; and Bid/Ask Spread is the bid/ask spread of the barrier option set as 2\% of the payout jump at expiration. All approximation errors are much smaller than the bid/ask spread.}
\label{tabitmbarsrf0}
\end{center}

As with earlier examples, the approximation is less accurate when risk neutral drift is significant. Table \ref{tabitmbarsrf5} shows results for the same market data except with a risk neutral drift of -5\%. Approximation errors are sometimes half the bid/ask spread but generally less.

\begin{center}
\begin{tabular}{| c | c | c | c | c | c | c | c | c |}
\hline
\textbf{Call/} & \textbf{Strike} & \textbf{Barrier} & \textbf{Model} & \textbf{Approx} & \textbf{Approx} & \textbf{Heston} & \textbf{BS} & \textbf{Bid/Ask} \\
\textbf{Put} & & & \textbf{Price} & \textbf{Price} & \textbf{Diff} & \textbf{Diff} & \textbf{Diff} & \textbf{Spread} \\ \hline
Call & 0.95 & 1.200 & 390.2 & 394.8 & -4.6 & -4.6 & +1.5 & 50 \\ \hline
Call & 1.00 & 1.200 & 124.7 & 128.4 & -3.7 & -3.8 & -19.4 & 40 \\ \hline
Call & 1.00 & 1.100 & 81.5 & 80.3 & +1.2 & -2.8 & -4.0 & 20 \\ \hline
Call & 1.00 & 1.050 & 24.0 & 18.1 & +5.9 & -1.3 & +8.0 & 10 \\ \hline
Call & 1.01 & 1.100 & 55.6 & 54.7 & +0.9 & -2.5 & -5.6 & 18 \\ \hline
Call & 1.03 & 1.100 & 22.5 & 22.0 & +0.5 & -1.8 & -5.1 & 14 \\ \hline
Call & 1.10 & 1.200 & 4.9 & 6.2 & -1.3 & -0.8 & -1.5 & 20 \\ \hline
Put  & 0.90 & 0.800 & 24.3 & 24.5 & -0.2 & -1.7 & -3.2 & 20 \\ \hline
Put  & 0.97 & 0.900 & 49.8 & 46.1 & +3.7 & -3.7 & -28.0 & 14 \\ \hline
Put  & 0.99 & 0.900 & 108.3 & 103.0 & +5.3 & -5.6 & -41.1 & 18 \\ \hline
Put  & 1.00 & 0.950 & 33.6 & 31.4 & +2.1 & -4.6 & +7.8 & 10 \\ \hline
Put  & 1.00 & 0.900 & 148.8 & 142.7 & +6.1 & -6.5 & -45.6 & 20 \\ \hline
Put  & 1.00 & 0.800 & 319.3 & 319.3 & -0.1 & -4.2 & -70.2 & 40 \\ \hline
Put  & 1.05 & 0.800 & 703.0 & 703.1 & -0.1 & -4.7 & -78.4 & 50 \\ \hline
\end{tabular}
\captionof{table}{Barrier price comparison for in-the-money barrier options when risk neutral drift $\mu=-5\%$. All prices and price differences are displayed in basis points. Approx Diff is the difference between the formal model price and the approximation; Heston Diff is the difference between the model price and the Heston price of the barrier option; BS Diff is the difference between the model price and the Black-Scholes price of the barrier option; and Bid/Ask Spread is the bid/ask spread of the barrier option set as 2\% of the payout jump at expiration. All approximation errors are much smaller than the bid/ask spread. Approximation errors are larger than the zero risk neutral drift case but still small compared to the bid/ask spread.}
\label{tabitmbarsrf5}
\end{center}

\section{Model Parameterization}

The pricing of barrier options in the approximation above depends on market implied volatilities and the model parameters; but a limited set of model parameters. In addition the approximation does not require a calibration of model parameters to implied volatilities: instead it requires a Heston model calibration which is much more efficient.

The subset of model parameters required by the approximation are $\beta$, $\gamma$, and $\rho_{cs} \epsilon$. Since the approximation does not depend on $\rho_{cs}$ or $\epsilon$ individually, but only on their product, we define $\xi = \rho_{cs} \epsilon$ as a separate parameter.

A derivatives market maker would be expected to mark those three parameters, which gives them the ability to calibrate (in an ad hoc way) to market prices of barrier options they see in the market. However, those marks can also be motivated by an estimation from historical dynamics, and this section explains how to estimate these parameters historically.

\subsection{Historical estimation of \boldmath{$\beta$}}

The mean reversion speed of volatility impacts the relative moves of at-the-money (ATM) implied volatilities of different tenors: when mean reversion is high the short term ATM implied volatilities move more than long dated ATM implied volatilities.

$\beta$ then can be estimated historically by looking at a regression of daily moves in ATM implied volatilities of different tenors. To do this we note that stochastic volatility does not affect the level of ATM implied volatility very much; so when analyzing moves in ATM volatilities we can simplify to a model with zero volatility of volatility. In this simplified model the ATM implied volatility for an expiration date $T$, $\sigma_I(T)$, is given by:

\begin{equation}
\sigma_I(T) = \sqrt{ \overline{v} + \frac{(v(0)-\overline{v})}{\beta T} \left( 1 - e^{-\beta T} \right) }
\end{equation}

Moves in $\sigma_I(T)$ come from moves in the instantaneous volatility $v(0)$, so we can write

\begin{equation}
d \sigma_I^2(T) = dv(0) \frac{\left( 1 - e^{-\beta T} \right)}{\beta T}
\end{equation}

If we then regress daily moves of implied volatilities of two different tenors $T_1$ and $T_2$, we find

\begin{equation}
d \sigma_I^2(T_2) = d \sigma_I^2(T_1) \frac{T_1}{T_2} \frac{ \left( 1 - e^{-\beta T_2} \right) }{ \left( 1 - e^{-\beta T_1} \right) }
\end{equation}

\begin{figure}[h]
\includegraphics[width=\textwidth]{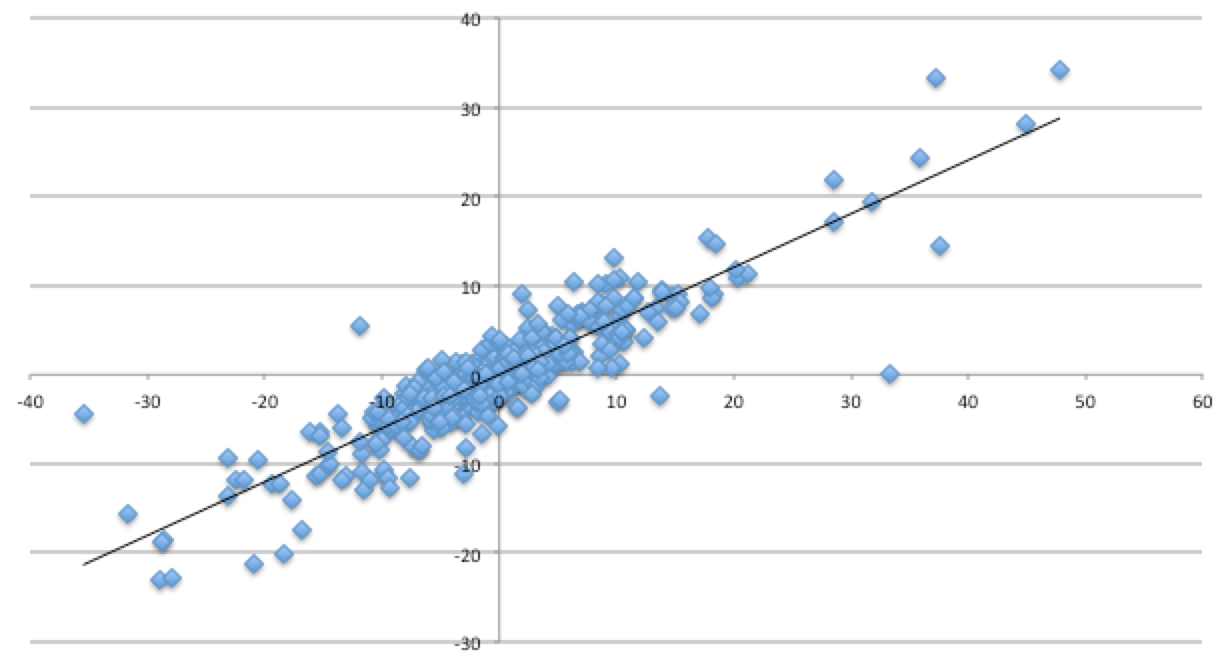}
\caption{A regression of daily moves in 1y ATM implied volatility squared against daily moves in 3m ATM implied volatility squared for EUR/USD from 1Jun2011 to 31May2013.}
\label{figatmreg}
\end{figure}

Figure \ref{figatmreg} shows a regression of daily moves in 1y ATM implied volatility squared vs daily moves in 3m ATM implied volatility squared for EUR/USD over a two year window from 1Jun2011 to 31May2013. The $R^2$ of a linear fit is 80\%, and the slope of 0.60 implies $\beta=1.63$, using $T_1=0.25$ and $T_2=1$. For G10 foreign exchange $\beta$ typically has values around 1-2.

\subsection{Historical estimation of \boldmath{$\gamma$}}

As mean reversion in volatility leads to a term structure of volatility of implied volatility, mean reversion in spot/volatility correlation leads to a term structure of volatility of risk reversal. To estimate $\gamma$, then, we need a way of approximately relating the risk reversal to model parameters, as we did with ATM implied volatility in terms of $\beta$ above.

\subsubsection{An Approximation for the Risk Reversal}
\label{secrr}

We start with the price premium of the risk reversal position from Equation \ref{eqrrprem}. We can use this to calculate the risk reversal implied volatility difference $RR(T)$:

\begin{equation}
v_{RR}(T) = ( V_C(T) + V_P(T) ) \frac{RR(T)}{2}
\end{equation}

where $V_C(T)$ and $V_P(T)$ are the vegas of the call and put options in the risk reversal position respectively. For a given delta, those are both roughly proportional to $\sqrt{T}$. So we can write

\begin{equation}
\label{eqrrapprox}
RR(T) \approx \frac{B \alpha}{\beta \sqrt{T}} \left( \overline{\rho} \, D_1(T) + (\rho(0)-\overline{\rho}) \, D_2(T) \right)
\end{equation}

where $B$ is another constant of proportionality that depends mostly on the delta of the risk reversal (we normally consider a delta of 0.25).

\begin{figure}[h]
\includegraphics[width=\textwidth]{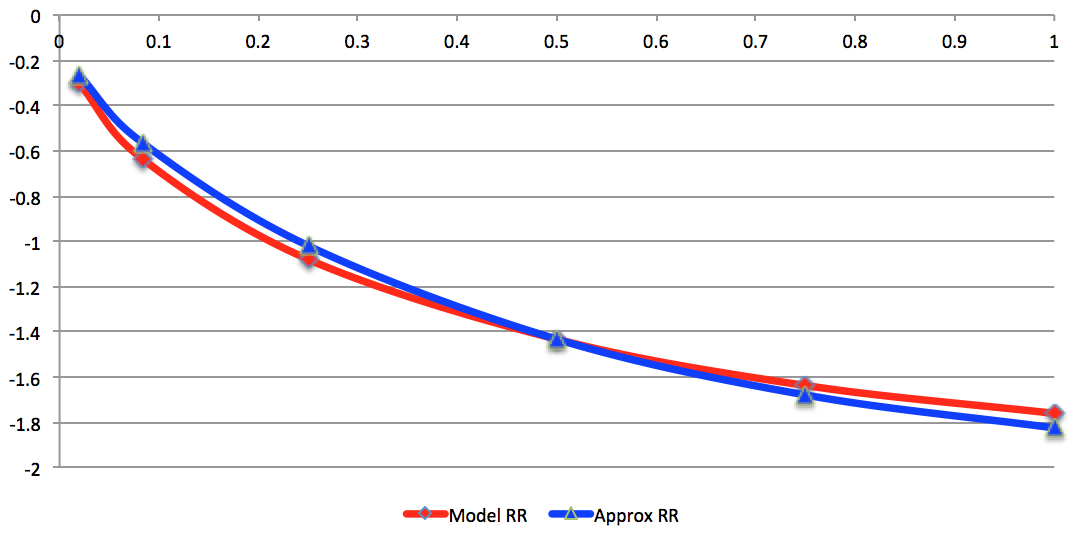}
\caption{SVSC and approximate risk reversal levels as a function of expiration tenor. The simple closed-form approximation does a reasonable job of matching the actual model risk reversal term structure after calibrating its scale to the risk reversal at $T=0.5$y.}
\label{figrr}
\end{figure}

We can test this against the value of the risk reversal in SVSC; the result is displayed in figure \ref{figrr}. This shows the risk reversal in SVSC against the approximation after fitting the value of $B$ to the point at $T=0.5$. In this example we used model parameters $\beta=2$, $\overline{v}=v(0)=0.01$, $\alpha=0.25$, $\gamma=4$, $\overline{\rho}=-0.6$, $\rho(0)=-0.25$, $\epsilon=10$, and $\rho_{cs}=0.7$, which are representative of the G10 foreign exchange markets. The approximation shows a close match to the true values even when the initial and average spot/vol correlation parameters are significantly different.

\subsubsection{Estimating \boldmath{$\gamma$} from risk reversal moves}

In this model moves in risk reversal are driven by moves in the instantaneous spot/volatility correlation $\rho(0)$, so we can write

\begin{equation}
d RR(T) \propto d \rho(0) \frac{D_2(T)}{\sqrt{T}}
\end{equation}

and so a regression of changes in risk reversal of two different tenors $T_1$ and $T_2$ is given by

\begin{equation}
d RR(T_2) = d RR(T_1) \frac{D_2(T_2)}{D_2(T_1)} \sqrt{\frac{T_1}{T_2}}
\end{equation}

$D_2(T)$ is a function of $\beta$ and $\gamma$; so given a value for $\beta$, we can use the slope of the regression to estimate $\gamma$.

\begin{figure}[h]
\includegraphics[width=\textwidth]{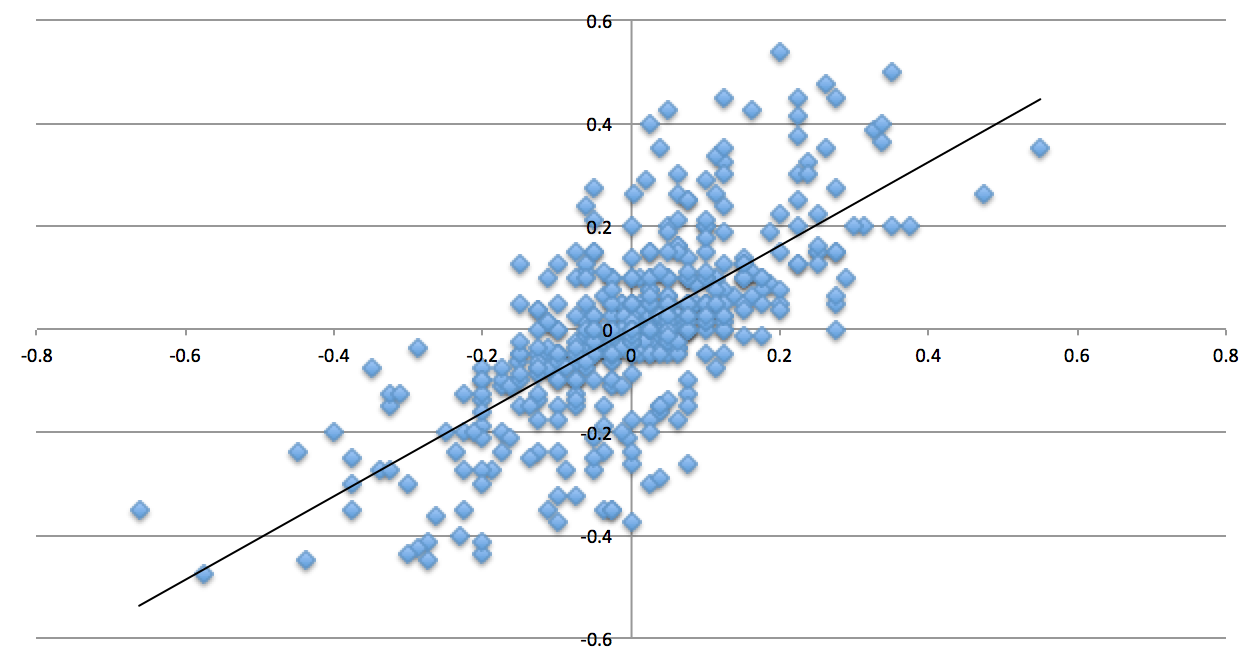}
\caption{A regression of daily moves in 1y risk reversal against daily moves in 3m risk reversal for EUR/USD from 1Jun2011 to 31May2013.}
\label{figrrreg}
\end{figure}

Figure \ref{figrrreg} shows a regression of daily moves in 1y risk reversal vs daily moves in 3m risk reversal for EUR/USD over a two year window from 1Jun2011 to 31May2013. The $R^2$ of a linear fit is 50\%, and the slope of 0.81 implies $\gamma=2.50$, using $T_1=0.25$, $T_2=1$, and $\beta=1.63$. For G10 foreign exchange $\gamma$ typically has values around 2-3.

\subsection{Historical estimation of \boldmath{$\xi$}}

Barrier pricing in the approximation depends only on the product $\xi=\rho_{cs} \epsilon$, not $\rho_{cs}$ or $\epsilon$ separately. $\xi$ effectively measures the key new feature of this model, the covariance between moves in spot and in the spot/volatility correlation. The risk reversal is the closest representation in market data to that spot/volatility correlation, and therefore to estimate $\xi$ we can look at covariance between moves in spot and moves in the risk reversal.

\subsubsection{The Risk Reversal \boldmath{$\beta$}}
\label{secrrbeta}

We define the``risk reversal $\beta$" as the expected move in risk reversal for a unit spot return to quantify that covariance. Since the expected move in $\rho$ given a spot return in this model is

\begin{equation}
\langle \frac{\partial \rho(t)}{\partial \ln(S)} \rangle = \rho_{cs} \epsilon \sqrt{1-\rho(t)^2}
\end{equation}

we can write the risk reversal $\beta$, $RR_\beta (T)$, in terms of $\xi$ as:

\begin{eqnarray}
RR_\beta (T) &=& \langle \frac{\partial RR(T)}{\partial \ln(S)} \rangle = \frac{\partial RR(T)}{\partial \rho(0)} \frac{\partial \rho(0)}{\partial \ln(S)} \nonumber \\
&\approx& \xi \sqrt{1-\rho^2(0)} D_2(T) \frac{B \alpha}{\beta \sqrt{T}}
\label{rrbetaapprox}
\end{eqnarray}

\begin{figure}[h]
\includegraphics[width=\textwidth]{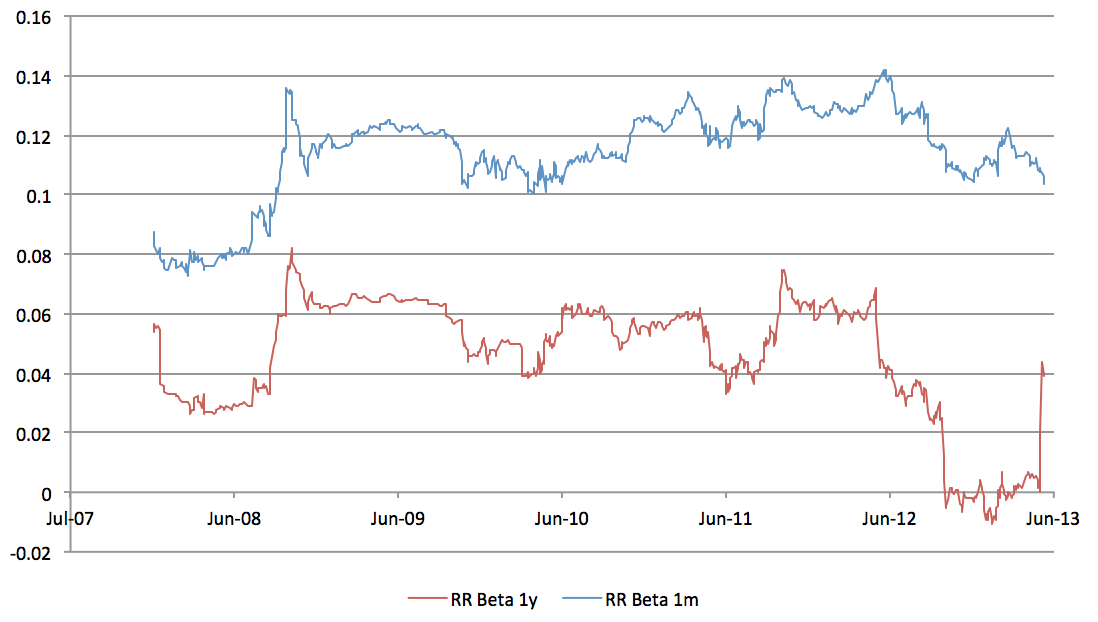}
\caption{Realized risk reversal betas for EUR/USD 1m and 1y, using London close market data.}
\label{figrrbetahistory}
\end{figure}

Figure \ref{figrrbetahistory} shows the historical 1m and 1y risk reversal $\beta$ for EUR/USD from July 2007 to July 2013, using a one year rolling window for a regression of daily change in risk reversal against spot return. Typical values range from 0.05 to 0.15.

\subsubsection{Estimating $\xi$ from the risk reversal \boldmath{$\beta$}}

The term structure of the risk reversal $\beta$ is determined by $D_2$ which is a function of the model parameters $\beta$ and $\gamma$, already determined in earlier sections. We will therefore use just one tenor's value of risk reversal $\beta$ to estimate $\xi$. For a given tenor $T$ we use the level of the market risk reversal $RR(T)$ to determine the proportionality constant $B \alpha$ and then the measured risk reversal $\beta$ to calculate $\xi$. In this analysis we assume that $\rho(0)=\overline{\rho}$, and on each day set $\overline{\rho}$ to the value from a Heston model calibration to implied volatilities of the same tenor $T$.

That is, we calculate $\xi$ as

\begin{equation}
\xi = \frac{RR_\beta (T)}{RR(T)} \frac{D_1(T)}{D_2(T)} \frac{\overline{\rho}}{\sqrt{1-\overline{\rho}^2}}
\end{equation}

On a given date we use $RR_\beta(T)$ from the rolling historical regression and $RR(T)$ from market data on that date. $D_1(T)$ and $D_2(T)$ are calculating using the values of $\beta$ and $\gamma$ estimated in sections above.

\begin{figure}[h]
\includegraphics[width=\textwidth]{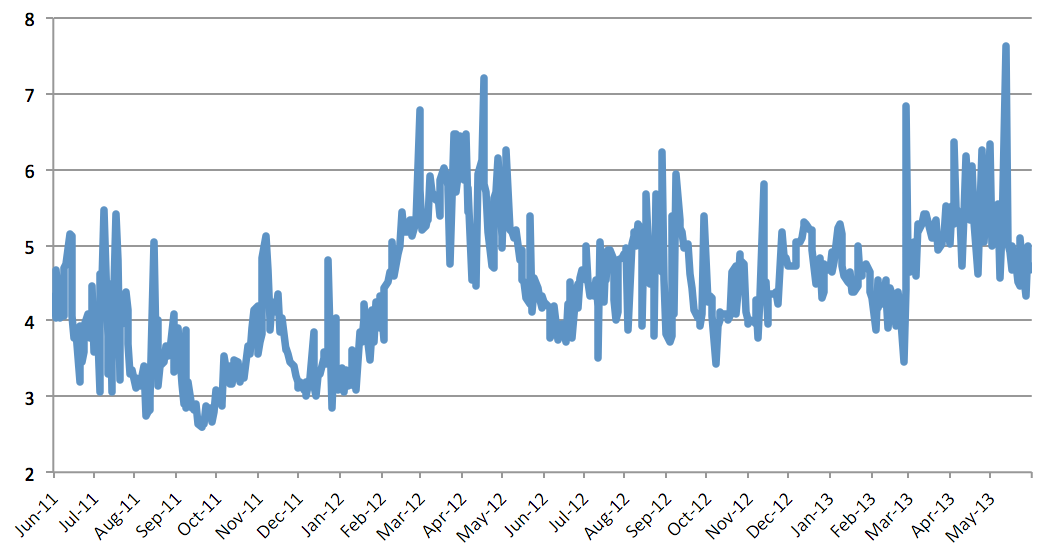}
\caption{Historical estimate of $\xi$ parameter for EUR/USD 1m-tenor market data in the 1Jun2011 to 31May2013 period.}
\label{figxicalc}
\end{figure}

Figure \ref{figxicalc} shows the historical estimate of $\xi$ for EUR/USD 1m-tenor data in the 1Jun2011 to 31May2013 period. Historical estimates of $\xi$ are typically in the 3-6 range.

\section{Conclusions and Future Improvements}

An important market dynamic for pricing barrier options is the covariance between spot and implied volatility skew, and the SVSC model developed in this paper captures that dynamic through a stochastic spot/volatility correlation that is itself correlated with spot.

We consider many variations of barrier options and one touches and show that the price impact of stochastic spot/volatility correlation is often comparable to or larger than the typical market bid/ask spreads for these products, and therefore is an important dynamic to consider - not surprisingly, as many market makers already use some flavor of LVSV model to approximately capture this effect. We discuss why the SVSC model is a more natural way of capturing the covariance between moves in spot and moves in implied volatility skew.

We also develop an approximation for barrier option pricing under the model that is both accurate and very fast, requiring only Heston model vanilla pricing and calibration. The approximation is accurate to a tolerance that is within the market bid/ask spread for barrier options in markets where the risk neutral drift is modest: a few percent or less.

A similar approximation for one touch prices was developed that is also accurate to a tolerance that is within the market bid/ask spread for one touches, again for markets where the risk neutral drift is modest.

In addition to market vanilla option prices the approximation requires three marked model parameters: $\beta$, the mean reversion strength of instantaneous variance; $\gamma$, the mean reversion strength of instantaneous spot/volatility correlation; and $\xi$, a parameter that controls the covariance between moves in spot and moves in the spot/volatility correlation. Those parameters can be marked to approximately calibrate pricing to market barrier options and one touches, or can be estimated historically. We demonstrate historical estimations of these three parameters.

Future work could include developing replications for barrier options and one touches that are better semi-static vega hedges in markets where risk neutral drift is significant.

\appendixpage
\appendix

\section{Dependence of Barrier Pricing on Correlation between Spot and Implied Volatility Skew}
\label{bardepcorr}

A key intuition about barrier option pricing is that their fair values depend mostly on what the model implies for covariance between spot and implied volatility skew, and not much on other aspects of the model. This is a good approximation for models where spot is diffusive - that is, does not allow for jumps in spot (though jumps in volatility are allowed).

This of course is not exact - it merely is an intuition pump to help modelers decide what aspects of the market dynamics to include in a model when pricing barrier options.

Consider barrier options which are out of the money when the barrier is hit: for example, a down-and-out call option. The boundary conditions that determine its value are the payout at expiration if the barrier was not hit (a call option payoff) and the value at the barrier level for times less than expiration (zero). As described in Section \ref{secotmbar} these boundary conditions can be replicated fairly accurately with a simple vanilla option portfolio: long a call option with the same strike and expiration date as the barrier option; and short a put option with the same expiration date as the barrier option, but a strike equal to the call option strike reflected through the barrier level.

If spot does not touch the barrier by expiration, the call option replicates the barrier option terminal payout. If spot is touched at some time before expiration, the two-vanilla replication roughly has a price of zero, as the long call and the short put are roughly the same amount out of the money and offset.

This is not an exact replication of course, even under Black-Scholes, especially when the risk-neutral drift is large. However, in practice the two-vanilla portfolio provides a good semi-static vega hedge for the barrier option - only semi-static because the hedge fails if the barrier is touched.

If the barrier is touched then the vanilla portfolio needs to be unwound, which is where model implications for dynamic hedging costs matter. At that point the long call and short put are roughly the same distance out of the money, and look like a risk reversal position. A risk reversal position has very little vega, so its value on unwind does not depend (much) on the model's prediction for at-the-money volatility when spot moves to the barrier. It also has very little sensitivity to the level of the implied volatility smile. To a good approximation the risk reversal position's value is linear in the implied volatility skew.

So the unwind cost - conditioned on hitting the barrier at some future time - depends on the model's prediction for the implied volatility skew at that point. And because the value of the risk reversal position is close to linear in implied volatility skew, only the expected value of the implied volatility skew, conditioned on hitting the barrier at that future time, is important - not the variance of that future implied volatility skew.

The expected value of the implied volatility skew, conditioned on spot moving to the barrier, is going to be mostly defined by the covariance between spot and implied volatility skew.

Therefore, the expected cost of unwinding the two-vanilla hedge depends mostly on that covariance, and not on other features of the model.

A similar argument be made about an up-and-out put option, where the replication is a put option with the same strike $K$ as the barrier option, and a call option with strike $B^2/K$.

Note that the assumption of no jumps in spot is particularly important for this explanation; if spot jumps far enough through the barrier then the two-vanilla replication will not look like a risk reversal position at unwind.

\section{Form of the Expected Future Vanna of a Vanilla Option}
\label{secvds}

In Section \ref{sechestoncorr} we use the fact that, in the Black-Scholes model, the vanna of a risk reversal position is approximately proportional to $\frac{(T-t)}{\sigma T}$, where $t$ is the future date, $T$ is the expiration date of the vanilla option, and $\sigma$ is the constant volatility. We derive that relationship here, and note that the approximation is valid when $\sigma \sqrt{T} \ll 1$.

The Black-Scholes vanna of a vanilla option has the form

\begin{equation}
V(t) = -d_2(t) \frac{\phi(d_1(t))}{\sigma}
\end{equation}

where $V(t)$ is the vanna of the option, taken here for simplicity to be the derivative of the vega with respect to the forward (rather than spot), and we assume zero discount rates to simplify the expressions (that simplification has no material impact on the result). $\phi$ is the probability density function of a standard normal variable:

\begin{equation}
\phi(x) = \frac{e^{-\frac{x^2}{2}}}{\sqrt{2 \pi}}
\end{equation}

$d_1$ and $d_2$ are the usual Black-Scholes expressions

\begin{eqnarray}
d_1(t) &=& \frac{\ln \left( \frac{F(t)}{K} \right) + \frac{\sigma^2 (T-t)}{2}}{\sigma \sqrt{T-t}} \\
d_2(t) &=& d_1(t) - \sigma \sqrt{T-t}
\end{eqnarray}

where $F(t)$ is the forward at time $t$ for settlement at time $T$, and $K$ is the strike price.

The strike term in $d_1(t)$ and $d_2(t)$ can be written in terms of the initial value $d_1(0)$:

\begin{eqnarray}
d_1(t) &=& \frac{x(t) + d_1(0) \sigma \sqrt{T} - \frac{\sigma^2 t}{2}}{\sigma \sqrt{T-t}} \\
d_2(t) &=& \frac{x(t) + d_1(0) \sigma \sqrt{T} - \sigma^2 T + \frac{\sigma^2 t}{2}}{\sigma \sqrt{T-t}}
\end{eqnarray}

where $x(t) = \ln(F(t)/F(0))$. $x(t)$ of course is normally distributed with mean $-\frac{\sigma^2 t}{2}$ and variance $\sigma^2 t$.

We can calculate the expected value of the future vanna $V(t)$ by integrating its expression over all values of $x(t)$:

\begin{equation}
\langle V(t) \rangle = \int_{x(t)=-\infty}^{\infty}{ V(x(t),t) \frac{\phi \left( \frac{x(t)+\sigma^2 t/2}{\sigma \sqrt{t}} \right)}{\sigma \sqrt{t}} dx(t) }
\end{equation}

Substituting the expression for $V$ in terms of $x(t)$, this integrates to

\begin{equation}
\langle V(t) \rangle = \phi \left( d_1(0)-\sigma \frac{t}{\sqrt{T}} \right) \frac{(T-t)}{\sigma T} \left( -d_2(0) + \sigma \frac{t}{\sqrt{T}} \right)
\end{equation}

Now consider the limit where $\sigma \sqrt{T} \ll 1$:

\begin{equation}
\langle V(t) \rangle = -d_2(0) \frac{\phi(d_1(0))}{\sigma} \frac{(T-t)}{T}
\end{equation}

Note that this equals the initial vanna of the option, but linearly decaying to zero as $t \rightarrow T$.

Finally: $d_1(0)$ and $d_2(0)$ are functions of the initial delta of the vanilla option (and equal in the limit $\sigma \sqrt{T} \ll 1$), so one can also write

\begin{equation}
\langle V(t) \rangle = A \frac{(T-t)}{\sigma T}
\end{equation}

where $A$ is a proportionality constant that only depends on the initial delta of the option.

\end{document}